\documentclass[conference]{IEEEtran}

\usepackage{cite}
\usepackage{amsmath,amssymb,amsfonts}
\usepackage{algorithmic}
\usepackage{graphicx}
\usepackage{textcomp}
\usepackage{xcolor}
\def\BibTeX{{\rm B\kern-.05em{\sc i\kern-.025em b}\kern-.08em
    T\kern-.1667em\lower.7ex\hbox{E}\kern-.125emX}}
\begin{document}

\title{Building Student Understanding of Quantum Information Science and Engineering through Projects on Applications to Medical Technologies}

\author{\IEEEauthorblockN{Jessica L. Rosenberg}
\IEEEauthorblockA{\textit{George Mason University}\\
Fairfax, VA, USA \\
https://orcid.org/0000-0002-5993-9069}
\and
\IEEEauthorblockN{Nancy Holincheck}
\IEEEauthorblockA{\textit{George Mason University}\\
Fairfax, VA, USA \\
https://orcid.org/0000-0001-6999-4072}
}

\maketitle

\begin{abstract}
Medical technologies, including quantum machine learning (QML) and quantum sensing, represent transformative tools for addressing some of the most pressing challenges in healthcare and drug discovery today. We discuss the ways that these topics engage the interest of high-school, undergraduate, and graduate students in understanding quantum information science and engineering. We describe how students built their understanding of these areas through a research project that allowed them to gain an understanding of the technology, its limitations, and the associated ethical considerations. We also consider the challenges of building this kind of work into the curriculum and of bringing students with interests in the biological and medical areas into quantum science and engineering. 
\end{abstract}


\section{Introduction}

The workforce in quantum science and engineering is growing and industry is noting the need for more people to be prepared in this field \cite{b1}. At the same time, quantum computing and sensing hold enormous potential for medical advancements. Drug discovery is a challenging and computer-intensive area that is currently being aided by the growth of machine learning and holds potential for the use of quantum computing to better model molecular interactions, analyze medical imaging and genomic data and more \cite{b2,b3}. Quantum sensing also holds potential for more sensitive imaging (MRI is already a quantum technology), detection of cancers and viruses, and more \cite{b4}. 

While medical technologies could see huge improvements from quantum computing and sensing, students interested in the biological sciences are the least likely science, technology, engineering, and mathematics (STEM) students to express an interest in quantum information science and engineering (QISE) \cite{b5}. This lack of interest is a potential barrier to the growth and development of the workforce in this interdisciplinary area, as it is a huge pool of untapped potential. At the same time, the biological sciences are significantly more diverse than the physical sciences and engineering. Engaging students from the biological sciences can help the field build a robust and diverse workforce as it grows. 

Providing students with the opportunity to learn about quantum science through research projects, particularly those focused on medical technologies, provides an opportunity to: (1) allow students to lead with their interests, thereby getting them more deeply engaged in the material, (2) provide a framework for talking about a topic that is otherwise very difficult to cover more than just superficially, (3) get students interested in QISE who might otherwise have considered QISE out of their area of interest, and (4) allow students to explore the ethical issues related to QISE as is important for developing an ethically aware workforce \cite{b6}.

\section{Data and Methods}
This study uses a qualitative approach \cite{b12} to understand how integration of applications of QISE, particularly those related to medical sensing and quantum computing for drug discovery, impact student interest and learning. We examine two populations of students for this work, high school students participating in the Pathways to Quantum Immersion Program and undergraduate and graduate students in the Ideas in Quantum Science and Technology course at George Mason University. 

\subsection{Pathways to Quantum Immersion Program}

The Pathways to Quantum Immersion Program (Pathways Program) is for rising high school seniors (referred to here as the ``high-school students''). The program includes 2 weeks of virtual learning on quantum concepts, applications, and careers followed by a week-long residential in-person experience that builds on the virtual component and includes visits to university and government research labs, industry facilities, and a site related to quantum policy. After the in-person program, students were given the opportunity to create a ``quantum vision poster'' for the Quantum World Congress. In 2022 the students were told ``You will identify a current issue or technology that can be addressed or improved through quantum innovations.  You will research it, envision what it might look like in 10 or more years, and describe the breakthroughs needed, pros \& cons, and obstacles.'' The prompts for 2023 and 2024 were similar. In the data analysis, posters that were based on medical applications of quantum information science were identified.

The high school students who participated in the Quantum Immersion Program submitted resumes as part of the application process. Forty-five resumes from the 57 program participants in 2022, 2023, and 2024 were examined to see if they had participated in biology or medical science related activities beyond basic coursework. Twelve of the students' resumes were not available for inclusion in the study. Some of these students entered the program through an adjacent program that did not collect resumes, while the rest were shared through an a cloud-based file storage and collaboration platform and were no longer accessible when we went to use them for research purposes. Student resumes were coded for student engagement in biology or medicine-related research and club activities prior to their participation in the program. 

Over the three years of the Pathways Program, small group interviews were conducted at the end of the in-person week. The transcripts of these interviews were coded for discussions related to biology or medical-related interests. Following an initial coding of resume and interview transcripts, quotes related to biology and medicine were compiled into a spreadsheet to facilitate review for patterns and themes.  

\subsection{Ideas in Quantum Science and Technology Course}

Ideas in Quantum Science and Technology (Ideas course) is a course at George Mason University that was taught in Spring 2024 and Spring 2025. The course has no pre-requisites and is open to undergraduate and graduate students. The course is designed to introduce students to quantum concepts and applications. Because there are no prerequisites, the course uses minimal math to explore these ideas, culminating in a project on an application of quantum science and engineering to a societal challenge. 

The student projects for the Ideas course were reviewed to identify projects on medical-related topics. The projects were then open-coded with particular attention to how students expressed their interests in medicine, perspectives on ethics, and connections to quantum technologies. 

The students in the Ideas course in 2025 participated in a focus group interview toward the end of the course. This interview was coded for discussions of interest in biology and medical-related subjects. The interview was also coded for students' motivations for studying quantum information science and for their views on the ethical implications when applying QISE to medical-related issues. 

Research questions:

\begin{enumerate}
    \item How do students connect interests in medicine with QISE? 
    \item How do research projects on quantum-related medical technologies support student learning and engagement in QISE?
\end{enumerate}

\section{Results}

\subsection{Quantum Immersion Program Students}

Most of the high school students who signed up for the Pathways Program had not participated in biology or medical activities according to their resumes. During the three years of the program, 12 of the 45 students (27\%) for whom we have resumes included something biology or medical-related. For two of those students, the only thing biology or medical related on their resume was membership in the Red Cross club. Other students were involved in the biology, neurology, or biomedical club. Three students had medical-related internships or research projects and one had a forensics internship.  These numbers of students involved in medical-related activities are reflected in the 6 of the 33 (18\%) quantum vision posters that were related to medical technologies. The poster topics ranged from imaging technologies using quantum sensing to drug discovery and cancer treatment to the development of brain-computer interfaces. 

For several with an interest in medicine, there was a sense that they were not sure how quantum would fit in, but Vivienne noted that by doing research for the vision poster, 

\begin{quote}
``It's completely shifted my perspective of quantum. It's hard to determine what I want to do, but when I was looking into stuff for the Quantum World Congress, it was, I used to be very focused on like a doctor route, right? Premed, stuff like that. But now I'm looking at the quantum technologies involved in medicine, and so it's more like, do I wanna work on that instead of working on people?"
\end{quote}

Chloe, who had a similar experience noted, ``I was looking at premed, but then seeing how quantum can influence technology I'd like to see how we can use quantum to impact healthcare,'' and another student noted, ``I started thinking about different career fields that would be affected by quantum technology. For example, as a doctor, I would have to use technology that's been developed, even if I wasn't involved in the creation process.''

Students with very little intervention and assistance were able to delve fairly deeply into their research topic of interest for their quantum vision posters. For example, one project on cancer treatment described how traditional computational methods struggle to distinguish cancer subtypes, while recent research has demonstrated the potential of quantum machine learning (QML) to tackle this problem, but still struggles to integrate large and diverse health datasets.  

QISE comes with key ethical issues that students need to consider. Others have advocated for weaving ethics into the QISE curriculum and we have included it as a part of the projects in which the students participate. While this was integrated into the project, the students received little direct instruction on ethical issues in this area. Nevertheless, students came up with a variety of ethical considerations, but did not have the space or background more than superficially. Some of the concerns that they cited included: data security including the protection of medical data from misuse and unauthorized access, the possibility that the technologies could exacerbate existing inequalities in health care, and risks from new sensors and quantum-enabled drug development. On the positive side they cited the medical benefits that could be accrued through these new technologies and the possibility of not needing animal testing and shortening the time needed for new developments. 

\subsection{Ideas in Quantum Science and Technology Students}

All 12 of the students (6 took the course in 2023 and 6 in 2024) who enrolled in Ideas in Quantum Science and Technology in 2024 and 2025 (the two years the course has been taught) were undergraduate students with physical science or engineering majors or graduate students in engineering or physics. The student pseudonyms, year they took the course, whether they were an undergraduate (UG) or PhD student, their major, and the titles of their projects are listed in Table 1. Even with no pre-requisites, the students drawn to a course on QISE were from the physical sciences and engineering. While they were all in these disciplines, one of the physics students, Orli, listed medicine as a future career of interest on the survey and one of the computer science students, Dylan, was considering a double major in bioengineering. 

\begin{table}[ht]
\caption{Ideas in Quantum Science and Technology Students}
\begin{center}
\begin{tabular}{|l|c|c|c|l|}
\hline
Pseudo. & Year & Deg. & Major & Project title\\
\hline
Yara & 2024 & UG & Info. Tech. & Quantum Sensors to\\
 &      &    &             & Solve Sleep \\
 Lance & 2024 & PhD & Elec. Eng.& Improving Emergency \\
 &      &    &             & Response with Smart \\
  &      &    &             & Traffic Management \\
 &      &    &             & and Quantum Sensors\\
 Charles & 2024 & PhD & Comp. Eng. & Quantum Magnetometers \\
 &      &    &             & for Seismic Activity \\ 
 &      &    &             & Monitoring\\
 Kris & 2024 & UG & Comp. Sci.  & Sustainable Climate: \\
 &      &    &             & Taking Action with \\
 &      &    &             & Renewable Energy\\
 Henry & 2024 & UG & Mech. Eng. & Quantum Sensing for \\
 &      &    &             & the Fishing Industry \\
 Garret & 2024 & UG & Comp.Sci & Qrypt Haven: \\
      &      &    &   \& Math           & Revolutionizing Quantum \\
      &      &    &             & Cryptography\\
\hline
 Orli & 2025 & UG & Physics & Implementing Quantum \\ 
 David & 2025 & PhD & Physics & Sensing Materials into \\
 Adam & 2025 & UG & Mech. Eng. & MRI Machines to \\
      &      &    &             & Improve Resolution \\
      &      &    &             & \\
 Dylan & 2025 & UG & Comp. Sci. & Quantum Drug\\
 Daniel & 2025 & UG & Physics\&Math &  Discovery\\
 Samir & 2025 & UG & Comp. Sci & \\
\hline

\end{tabular}
\label{tab1}
\end{center}
\end{table}

Despite the students' focus on the physical sciences and engineering, seven of the twelve (58\%) Ideas class students did projects related to medicine. One caveat with respect to these numbers is that in 2025 the projects were completed in groups. If the students had selected their own topic, three of the students in 2025 might have selected one that was not related to a medical application (their initial submission of topic ideas did not include medical applications), but 33\% selected this as an application area of interest.

Dylan who is considering adding a major in bioengineering noted, ``I've always been interested in the intersection between technology and biological systems and seeing kind of the quantum application in modeling different drugs and how we can actually affect biological systems."
For Daniel and Adam, the project had connections to their academic interests. Daniel had previously worked on applications of quantum algorithms to drug discovery for a project with a professor. Adam's senior capstone project was on the application of nanotechnology and materials research to MRI and other medical technologies. In all cases (the medical-related projects as well as those on other topics), students gained the ability to talk about the technologies as seen in this excerpt from the project of Dylan, Daniel, and Samir:

\begin{quote}
``The promise of Quantum Drug Discovery (QDD) lies in its potential to accelerate and improve early-stage drug development through more accurate predictions of molecular behavior, binding affinities, and reaction mechanisms. For instance, quantum algorithms such as the Variational Quantum Eigensolver (VQE) and Quantum Phase Estimation (QPE) are being explored for their ability to calculate molecular ground-state energies — a key quantity in determining drug-ligand binding \cite{b11}. These methods, while still in early stages of development, are already demonstrating competitive accuracy on small molecules compared to classical computational chemistry tools."
\end{quote}

While they were able to talk about the technologies, they also described that quantum computing for drug discovery faces substantial challenges, ``Current quantum hardware, known as Noisy Intermediate-Scale Quantum (NISQ) devices, are limited by short coherence times and error rates that constrain their usefulness for large, biologically relevant systems."

All of the students in the Ideas course were able to talk more deeply and with more nuance about the ethical issues confronting quantum technologies than the high-school students. However, the discussions in Spring 2025 were deeper and more nuanced still as these students were provided with more references and spent more time discussing the issues in class. Likewise the students in 2024 spent more time discussing the issues in class than the high-school students who, other than a brief discussion during the in-person portion of the program did not have an opportunity to discuss the issues with others. The Ideas course students also had more space in their papers to talk about the ethical issues than the Pathways students did on their posters. For example, when one group of Ideas students talked about the issue of data privacy they went beyond just the generic issue of misuse and unauthorized access to talk about how quantum-related technologies (particularly with respect to brain sensing) could produce ``much larger and more detailed datasets" and could ``uncover incidental findings" that raise issues with respect to ``informed consent, patient counseling, and follow-up care."

\section{Discussion and Conclusions}

While some students came to the Pathways Program and the Ideas course with an interest in biology or medicine, neither program drew students for whom medicine is a primary interest. Quantum science is an interdisciplinary field that should be well positioned to draw students with a variety of interests and bachelor's degrees in the biological sciences make up $\sim30\%$ of STEM degrees \cite{b12}. In order to grow the field and bring in perspectives of students interested in the biological or medical sciences, different forms of recruitment are going to be needed. Just because these opportunities are open to students with a range of backgrounds and interests does not mean that they will sign up without additional encouragement and help understanding why they might be interested in quantum.  

Even though biology and medicine were not the primary interest of most of the students, many chose to explore these areas when they were asked to consider the impact of QISE on a societal challenge (Ideas course) or to envision the impact of QISE on a current issue or technology (Pathways program). Through these projects, they found connections to topics that they were interested in, particularly with respect to applying quantum science to a societal challenge. Through independent exploration, they were able to get to a point where they could identify how quantum sensing or computing could be applied to the challenge (e.g., drug discovery or brain imaging) and use that as a mechanism for building a much deeper understanding of the technology. These efforts also allowed the students to explore the advantages and limitations of quantum sensing and computing for these applications. In general, the students went much further in the development of their understanding of the technologies with much less direct guidance than they likely would have with a very directed class activity or assignment. 

As much as the students were able to learn through Pathways or the Ideas course, there was a limit to the understanding they could develop in a conceptual course with no math prerequisites. If this course is intended to lead into further coursework in QISE, it will be important to examine whether it helps the students prepare for the next level of the curriculum where they will need to bring in the math. 

By asking students to engage with a societal challenge, a current issue, or a QISE technology, students had a platform upon which to consider the ethical challenges of applying quantum technologies. While some of the answers were fairly superficial, when pressed do delve more deeply into the issues and provided with some relevant reading material as they were for the Ideas course in 2025, they were able to discuss the ideas in insightful and nuanced ways. Access to relevant resources and opportunities for discussion are important for building deeper understanding of the ethical issues among the students. 

Examining applications of QISE, particularly biomedical applications, allowed students to think about the subject in a new way and to consider how it can help or harm society. Applications and ethics are topics that are often missing or limited in scope within STEM curricula. QISE applications could also be a way to try to engage more students who are interested in the biological sciences. However, a student interested in drug discovery will have to take significant amounts of mathematics and computer science in addition to QISE to continue in that field. A student interested in brain imaging would not need significant QISE knowledge to make use of such systems but would need significant physics and/or engineering experience to develop the systems. Because this kind of project would only be a first step for a student interested in continuing with QISE, it will only be an effective recruitment mechanism if it occurs early, likely in high school and with students who do not have to self-select into a quantum program. 

\section*{Acknowledgment}
Thank you to Melinda Ryan, George Thomas, and Kieran Collinson for their efforts supporting students in all aspects of the Pathways program. We thank George Thomas particularly for making it possible for the students to present their projects at the Quantum World Congress.

\end{document}